# All Inkjet-Printed Graphene-Silver Composite Inks for Highly Conductive Wearable E-Textiles Applications

**Nazmul Karim,[1*] Shaila Afroj,[1,2] Sirui Tan,[3] Kostya S Novoselov[1,2] and Stephen G. Yeates[4]**

[1]National Graphene Institute, The University of Manchester, Manchester, M13 9PL, UK,

[2]School of Physics and Astronomy, The University of Manchester, Manchester, M13 9PL, UK,

[3]School of Materials, The University of Manchester, Manchester, M13 9PL, UK

[4]School of Chemistry, The University of Manchester, Manchester, M13 9PL, UK

*Corresponding E-Mail (Authors): mdnazmul.karim@manchester.ac.uk

**Abstract**

*Inkjet-printed wearable electronic textiles (e-textiles) are considered to be very promising due to excellent processing and environmental benefits offered by digital fabrication technique. Inkjet-printing of conductive metallic inks such as silver (Ag) nanoparticles (NPs) are well-established and that of graphene-based inks is of great interest due to multi-functional properties of graphene-based materials. However, poor ink stability at higher graphene concentration and the cost associated with the higher Ag loading in metal inks have limited their wider use. Moreover, graphene-based e-textiles reported so far are mainly based on graphene derivatives such as graphene oxide (GO) or reduced graphene oxide (rGO), which suffers from poor electrical conductivity. Here we report inkjet printing of highly conductive and cost-effective graphene/Ag composite inks for wearable e-textiles applications. The composite inks were formulated, characterised and inkjet-printed onto PEL paper first and then sintered at 150 °C for 1 hr. The sheet resistance of the printed patterns is found to be in the range of ~0.08 – 4.74 Ω/sq depending on the number of print layers and the graphene/Ag ratio in the formulation. The optimised composite ink was then successfully printed onto surface pre-treated (by inkjet printing) cotton fabrics in order to produce all-inkjet-printed highly conductive and cost-effective electronic textiles.*

Printed flexible electronics have demonstrated great potential for further growth with the applications in health diagnostics[1,2], energy storage[3], food security[4], touch screens[5], electronic paper[6], sensors[7,8], radio frequency tags[9] and electronic textiles[10]. Among them, wearable electronic textiles (e-textiles) that contain electronics and interconnections woven/knitted or printed/coated into or onto textiles[11], are of great interest as they provide better comfortability, durability and lighter weight as well as maintaining the desirable electrical property. Inkjet printing is one of the most promising techniques for the fabrication of flexible and wearable electronics, due to a number of advantages over conventional manufacturing techniques such as digital and additive patterning, reduction in material waste, deposition of a controlled quantity of materials and compatibility with various substrates[12,13]. Currently silver nanoparticles (NPs) based inkjet inks are the most reported and utilised conductive inks due to their excellent electrical conductivity and strong antioxidant characteristics[14-16]. However, a higher concentration of NPs and higher sintering temperatures are required in order to obtain a continuous metallic phase, with numerous percolation paths between the metal particles within the printed pattern[17,18]. Moreover, the higher price of silver increases the processing cost[15]. In addition, some of the printed substrates are also sensitive to higher temperature, which limited the choice of the substrates to be printed[14].

Graphene, a single atom thick two-dimensional closely packed honeycomb lattice of $sp^2$ carbon allotropes, has been the focus of extensive investigations in recent years due to its unique physical and chemical properties[19,20]. Inkjet printing of graphene-based inks is an encouraging research approach in the field of printed electronics as both the benefits of inkjet printing[12] and the extra-ordinary electronic[21], optical and mechanical properties[22,23] of graphene can be exploited[24]. Several studies[10,25-27] have reported inkjet printing of reduced graphene oxide (rGO) as a popular choice to fabricate flexible devices, as it is readily dispersible in water, and higher volume can be produced at lower cost[28]. Our recent studies have shown promise of using graphene derivatives such as GO or rGO for wearable e-textiles applications[3,10,29-31]. However, the large number of unreduced oxygen-containing functional groups and inter-sheet junctions between the graphene flakes limit the conductivity achieved with rGO[32]. In order to overcome such limitations associated with rGO inks, pristine graphene inks were developed and used for printing. However, a number of limitations still exist with such inks and their fabrication process developed so far: such as lower graphene concentration and residual solvent in the printed electronics[33,34], time consuming process and toxic solvents[24], repeated and complicated process[12], a higher annealing temperature and a number of inkjet layers needed to achieve the desired conductivity[35].

In order to overcome these challenges, here we propose the formulation of highly conductive silver nanoparticles decorated graphene-based composite inkjet ink that can achieve desired electrical conductivity with few layers of printing. Previous studies[15,17,36-39] reported composite formulations containing metal nanoparticles such as Ag and Cu, and carbon-based materials such as graphene and carbon nanotubes (CNT), however limited to lower concentration and not demonstrated for printing onto challenging substrates like textiles for wearable applications. Here we report inkjet printing of highly conductive graphene/silver nanoparticles composite inks for wearable e-textiles applications. We optimise the formulation using a cube film applicator first and then formulated a range of composited inkjet inks. Finally the optimised graphene/silver composite ink was printed on surface pre-treated (by inkjet) textiles for highly conductive wearable e-textiles applications.

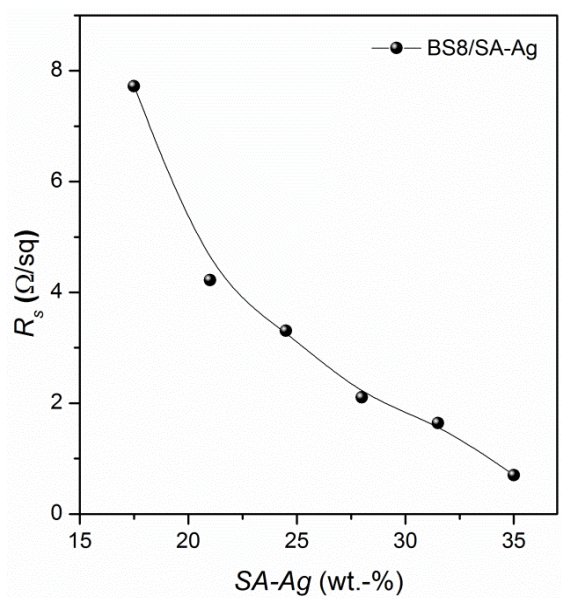

**Figure 1** Sheet resistances of the conductive films drawn on a glass substrate with SA-Ag/BS8 composites using a cube film applicator.

**Results and Discussion**

**Optimisation of Composite Formulations**

A range of composite formulations were prepared using a silver nanoparticle-based ink (SA-Ag, ~30-35 wt.-%) and a highly concentrated graphene dispersion (BS8, ~8 wt.-%). A simple triple reservoir cube film applicator was used to produce conductive films (~90 μm thick) of composites formulations on a glass substrate. The formulations were then optimised by measuring the sheet resistance of such conductive films. The conductive film with BS8 ink on glass substrate provides higher sheet resistance (~32 kΩ/sq). As expected, with the addition of SA-Ag to the composite ink formulation, the sheet resistance of the conductive film decreases with the increase of wt.-% of highly conductive SA-Ag (as received) nanoparticles (~0.7 Ω/sq), Figure 1. However, a minimum of 17.5 wt.-% of SA-Ag (as received) is required in the composite formulations in order to form a continuous conductive film (~ 8 Ω/Sq) on a glass substrate.

In order to demonstrate the effect of graphene flakes on the conductivity of composite formulations, SA-Ag inks were also diluted with TEGMME and 1 wt.-% PVP in TEGMME. Again the conductive patterns were drawn using a cube film applicator and sintered at 150 °C for 1hr. The results are compared in Table 1 with that of SA-Ag/BS8 to demonstrate the contribution of graphene to the conductivity of composite ink formulations. Table 1 shows that SA-Ag inks diluted with TEGMME provides higher sheet resistance (~20.45 Ω/Sq) than that of other diluted inks. With the addition of PVP, the sheet resistance of conductive films decreases, as PVP stabilises nanoparticles dispersions with higher metal loadings to provide higher conductivity[14]. PVP is a linear homopolymer of N-vinylpyrrolidone, and soluble in water and other polar solvents. It forms a coating on graphene flake surfaces to prevent aggregation in aqueous dispersion[40]. The presence of graphene flakes in the SA-Ag/BS8 composite increases the electrical conductivity, as the conductive film drawn with such ink imparted lowest sheet resistance (~7.73 Ω/sq) out of three formulations presented in Table 1. The density of Ag is almost 10 times higher than that of graphene[41,42], therefore in terms of volume the SA-Ag/BS8 composite formulation is mostly dominated by BS8 (70.57 vol.-%) with less Ag (either vol.-% or wt.-%) content than the other formulations. Graphene flakes provide an effective pathway in both the length and width directions of the conductive film to avoid negative influence of the void[15]. Moreover, graphene has very high surface area with theoretical value ~2600 $m^2 g^{-1}$, which causes spread out of the transparent graphene in all directions of the film to create a bridge between the aggregates of Ag NPs. Thus good contact between graphene and silver is achieved to provide better conductivity.

**Table 1** Sheet resistances of conductive films drawn on glass substrate produced by SA-Ag inks blended with BS8, TEGMME and 1% PVP in TEGMME. (density, $\rho$ of graphene[42] ~1 g/cm³ and bulk silver(Ag)[41] ~ 10.49 g/cm³).

| Formulations | SA-Ag/BS8 | SA-Ag/1% PVP in TEGMME | SA-Ag/ TEGMME |
|---|---|---|---|
| As received (wt.-%) | 50/50 | 50/50 | 50/50 |
| wt.-% | 17.5/4 | 17.5/0.5 | 17.5/0 |
| %Bulk (wt/wt) | 81.40/18.60 | 97.22/2.78 | 100/0 |
| %Bulk (vol/vol) | 29.43/70.57 | 80.02/19.98 | 100/0 |
| Sheet Resistance (Ω/Sq) | ~7.73 | ~9.34 | ~20.45 |

**Inkjet Printing of Composite Inks**

A Dimatix Materials Printer (DMP) was used for inkjet printing of conductive inks onto a PEL paper first. Therefore, the surface tension of BS8 graphene dispersions was adjusted to ~34.5 mN/m by adding 1 wt.-% non-ionic surfactant (Triton X-100) to make it inkjettable in DMP. However, the higher voltage was required to jet ink droplets through the nozzles and chaotic jets were observed after printing three layer of a 1 cm$^2$ conductive pattern with formulated inks, even after adding a stabiliser (PVP). This may be due to the van der Waals forces between highly concentrated graphene sheets with a higher aspect ratio, which causes them to stick together to form large bundles or ropes and clogs the nozzles of the inkjet print head[14,43]. The sheet resistance of the printed patterns is found to be ~49.26, ~40.99 and ~28.13 Ω/sq for 1, 2 and 3 layers, respectively, which are lower than those achieved with inkjet-printed pristine graphene in several previous studies reported in this review[44]. Moreover, BS8 provides lower sheet resistance on surface pre-treated PEL paper than glass, may be due to the better overlapping and connections between individual flakes. Please note, the thickness of the conductive film on paper cannot be measured accurately due to the roughness of the substrate.

In contrast, the viscosity and surface tension of SA-Ag inks are reported as 10-18 cps and 30-40 mN/m, respectively, as per the technical information supplied by Sigma-Aldrich, which are within the DMP printing range. The ink was jetted effectively through all the inkjet nozzles and 16 jets were working well at 25 V. The sheet resistance of the conductive pattern printed with SA-Ag inks are 0.07, 0.04, 0.04, 0.03 and 0.03 Ω/Sq for 1L, 2L, 3L, 4L and 5L, respectively (Figure 2a). It is assumed that some of the SA-Ag nanoparticles were diffused into PEL paper in the first layer of printing and then created a conductive film on the surface. In the subsequent layers of inkjet printing, Ag nanoparticles were deposited on that film with less reduction in the sheet resistances, Figure 2a. Again it was impossible to measure the accurate thickness of the conductive film on paper; however, almost the same amount of SA-Ag inks was printed during each layers of inkjet printing due to the same drop size and drop spacing.

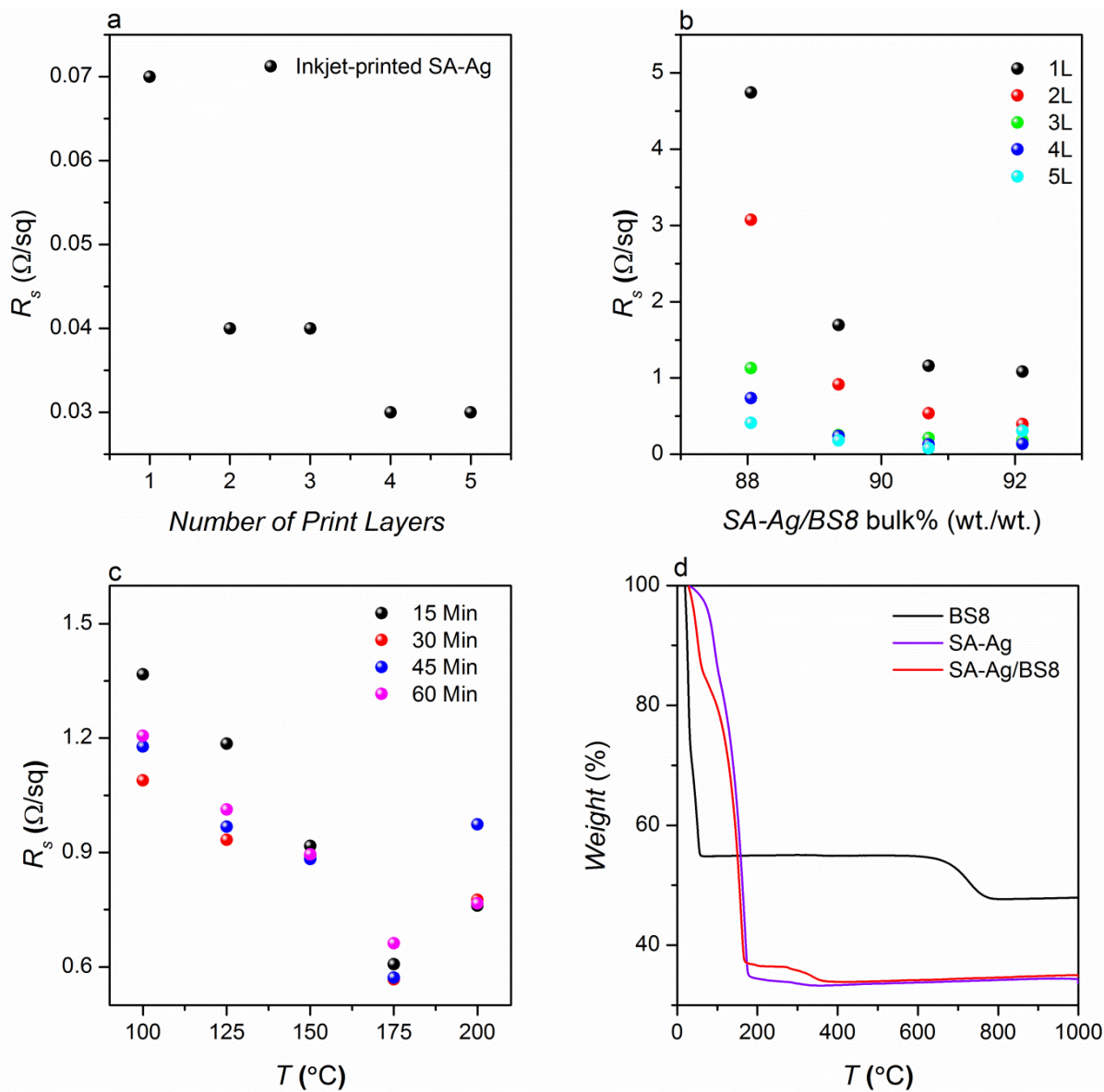

**Figure 2** a) The change of sheet resistance of inkjet-printed SA-Ag with number of layers; b) Effect of SA-Ag/BS8 (vol.-%) ratio and the number of layers on the sheet resistances of the inkjet-printed conductive patterns; c) Effect of annealing temperature and time on the sheet resistance of inkjet-printed conductive patterns printed with composite Ink C; and d) TGA diagrams of SA-Ag, BS8 and composite Ink C (SA-Ag/BS8).

We then formulated a range of silver decorated graphene-based composites inks (Table 2) on the basis of the results obtained from composite ink optimisation. Ink A jetted initially at a higher voltage 40 V with a highly chaotic and irreproducible droplets and jetting was stopped after a few minutes of printing. This is attributed to the poor solubility of highly concentrated (8 wt.-%) BS8 dispersion, due to the attractive van der Waals forces between graphene sheets, which causes re-aggregation even after the initial process of exfoliation and dispersions[45]. In order to improve the stability of the formulated ink, a polymer stabiliser PVP (dissolved as 1

wt.-% in TEGMME) was then added to Ink A in different quantities in order to prepare a range of inks, Table 2. The addition of PVP provided improved the jettability of ink droplets *via* DMP inkjet printer and uninterrupted printing of a few layers of composite inks. Firstly, TEGMME (~31.4 mN/m) has lower surface tensions, which helps to reduce the surface tension of composite ink formulations and keep it within the range of DMP inkjet printer. Secondly, PVP improves the ink stability by forming a coating on graphene flake surfaces to prevent aggregation in aqueous dispersion[40]. In addition, PVP is the most commonly used stabiliser for metal NPs-based inks in various solvents, which interact with metal NPs through the oxygen atom of the carbonyl group[46]. However, the presence of a stabiliser in a higher quantity could have a negative effect on the conductivity, as it acts as an insulating organic layer on NPs surfaces that prevents close electrical contacts between NPs[14].

**Table 2** The composition of graphene-based composite inks

|  | Solid (wt.-%) | | | % Bulk (wt/wt) | | | % Bulk (Vol/vol) | | |
|---|---|---|---|---|---|---|---|---|---|
|  | BS8 | SA-Ag | 1% PVP (TEGMME) | BS8 | SA-Ag | 1% PVP (TEGMME) | BS8 | SA-Ag | 1% PVP (TEGMME) |
| Ink A | 3.2 | 21 | 0 | 13.22 | 86.78 | 0 | 61.52 | 38.48 | 0 |
| Ink B | 2.8 | 21 | 0.05 | 11.74 | 88.05 | 0.002 | 57.81 | 41.33 | 0.86 |
| Ink C | 2.4 | 21 | 0.1 | 10.21 | 89.36 | 0.004 | 53.51 | 44.63 | 1.86 |
| Ink D | 2 | 21 | 0.15 | 8.64 | 90.71 | 0.006 | 48.46 | 48.51 | 3.03 |
| Ink E | 1.6 | 21 | 020 | 7.02 | 92.11 | 0.009 | 42.46 | 53.12 | 4.42 |

Figure 2b shows the effect of bulk wt.-% of SA-Ag NPs and the number of printed layers on the sheet resistance of the inkjet-printed conductive patterns with composite inks on PEL paper. As expected, the sheet resistance of the inkjet-printed patterns decreases with the increase of the bulk wt.-% of SA-Ag in the formulation. The highly conductive metallic SA-Ag NP ink provides numerous percolation paths between the metal particles within the printed patterns. In addition, the contact resistance between the graphene sheets is found to be one of the factors impeding the achievement of higher conductivity[17]. The coating of Ag NPs onto graphene sheets increases the contact interfaces and reduces the contact resistance between graphene sheets; thus improves the conductivity of the film[47]. Moreover, graphene also provides effective electrical networks between the silver flakes[36], by reducing the negative influence of the voids present in the printed silver film[17]. The sheet resistance of a single layer (1L) inkjet-printed conductive pattern with Ink B is found to be ~4.74 Ω/Sq, which is reduced approximately by 3 times to ~1.70 Ω/Sq when the bulk wt.-% and vol.-% of SA-Ag increases to 44.63% and 89.36%, respectively, in the formulation, Ink C (Table 2 and Figure 2b). In addition, the number of inkjet-printed layers is found to have a significant effect on the sheet resistance, as it decreases with the increase in the number of print layers. The sheet resistances achieved onto PEL papers by printing 1L and 5L of Ink B are found to be 4.74 and 0.41 Ω/sq, respectively, which indicates an approximately 50% reduction of the sheet resistance for each inkjet-printing layer of composite Ink B. Considering the trade-off between the efficient inkjet printing and electrical performance of the printed patterns on PEL paper, the composite Ink C was used for the subsequent process.

The presence of stabilising agents, organic solvents and other polymeric additive in conductive ink formulation hinders the effectiveness of electrical properties. Therefore, it is desirable to remove them after inkjet printing by sintering (most commonly "heating") in order to form a continuous electrically conductive film. The temperature required for sintering varies from ink to ink due to the different boiling points of the organic solvents, dissociation temperatures of the salts, the diameter of the NPs and graphene flakes. Figure 2c shows the effect of the annealing temperature and time on the sheet resistance for inkjet printing of 1L of composite Ink C. The annealing temperature and time are found to have a significant effect on the electrical properties of the printed pattern. As shown in Figure 2c, the sheet resistance decreases rapidly with the increase of annealing temperature up to 175 °C. This may be attributed to the thermal degradation behaviour of the composite ink, Figure 2d. The sheet resistance decreases from ~1.37 Ω/Sq to ~0.76 Ω/Sq as the annealing temperature increased from 100 °C to 175 °C, may be due to the thermal decomposition of PVP and the coalescence of the Ag NPs and good inter-connection between the graphene sheets in at higher temperature. This will enable them to form a continuous conductive path and provides good electric charge transports within the inkjet-printed track[38]. Similarly, the sheet resistance decreases with the increase of the annealing time up to 30 min, may be due to the increased number of aromatic compounds in the graphene network with a longer annealing time; however the concentration remains unchanged after 30 min[35].

As illustrated in Figure 2d, the TGA graph of SA-Ag indicates the volatilisation of solvents, stabiliser and other polymeric additives below 180 °C, leaving approximately 35 wt.-% solid silver, which is similar to the ~30-35 wt.-% solid content reported in Sigma-Aldrich product information. In contrast, the first decomposition stage is observed at a lower temperature, from 20 °C to 70 °C, for BS8 graphene dispersion, where 45 wt.-% of BS8 constituents are removed. The second stage decomposition for the same dispersion occurs between 595 °C to 845 °C, to remove the 7 wt.-% constituent, which might be attributed to the air oxidation of graphene above 400 °C[38]. The thermal decomposition of composite ink (Ink C) is found to be almost similar to that of SA-Ag ink; however, the second stage decomposition for that ink is observed from 265 °C to 380 °C with a slight decrease in weight (~2.45 wt.-%).

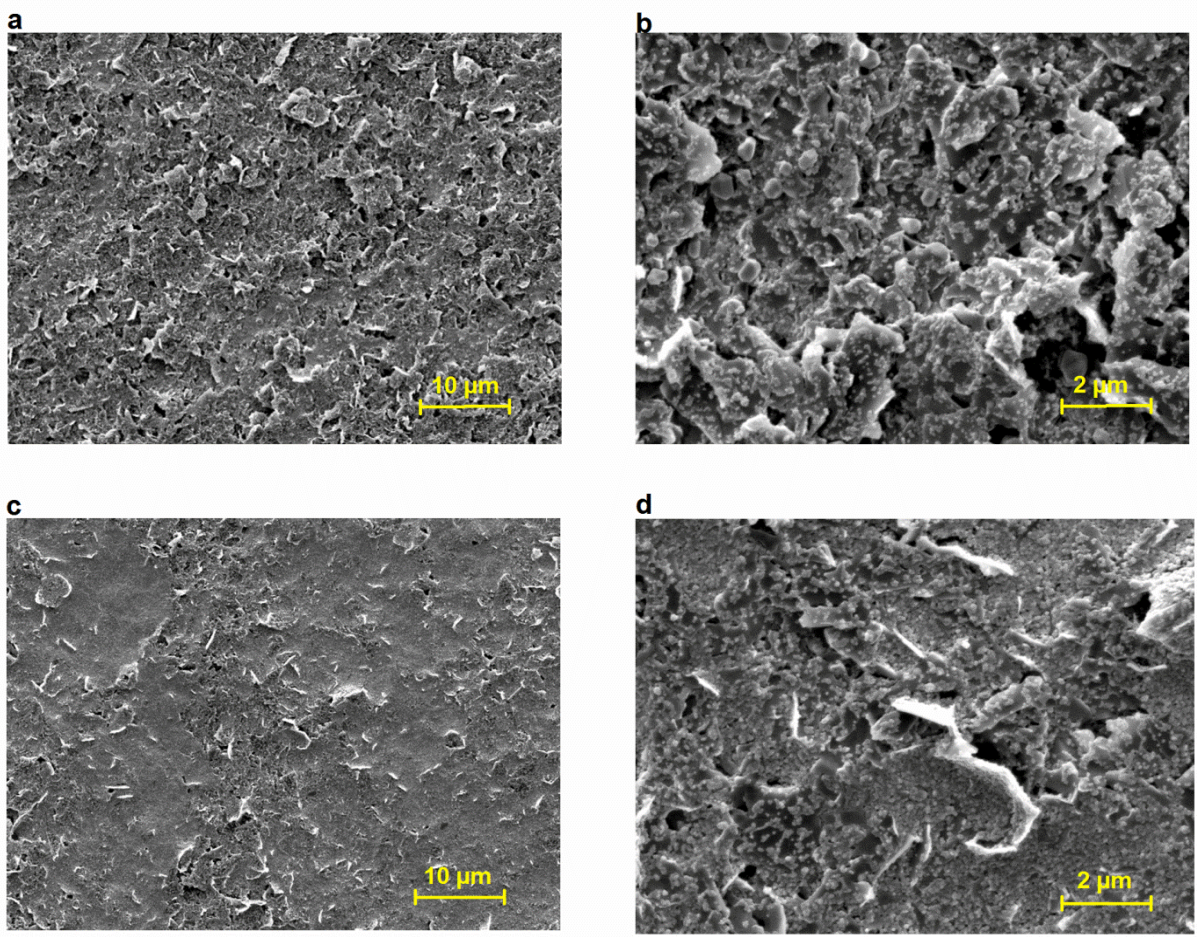

**Figure 3** SEM images of inkjet-printed composite Ink C onto PEL paper (a) 1L at 2000X; (b) 1L at 10000X; (c) 5L at 2000X and (d) 5L at 10000X

SEM was used in order to analyse the surface topography of the printed pattern on PEL paper. Figure 3 (a-d) shows the SEM images of inkjet-printed conductive pattern (1L and 5L) with composite Ink C. The images show the deposition of Ag NPs onto micrometre-sized graphene flakes and thicker conductive film with an increased number of printed layers, which resulted in lower sheet resistance. In addition, the higher loading amount of graphene reduces the voids present in Ag film by creating an effective conductive path. Moreover, the coalescence of Ag NPs is observed after sintering at 150 °C for 1 hr, which is increased with the increase of the number of printed layers.

**All Inkjet-Printed Wearable E-Textiles**

Inkjet printing of electrically conductive pathway on textiles is of interest, as inkjet offers a number of advantages over conventional manufacturing techniques, such as high resolution printed circuit, deposition of controlled amount of conductive materials at precise locations, reduction of materials waste and usage of natural resources such as waters, gas and electricity[12]. However, inkjet printing of low viscosity inkjet inks onto a rough and porous textile surface is of great challenge, due to the orientation of fibres or yarns[48] and the change of fibre morphology constantly due to the exchange of water molecules with surroundings[9]. In order to address these challenges, in our previous study[10], we reported synthesis of a novel surface pre-treat (NP1) and inkjet printing of 12 layers NP1 onto rough and porous textiles with subsequent printing of rGO on surface pre-treated textiles in order to produce continuous conductive pathway onto textiles. However, rGO surfers from poor electrical conductivity due to the partial restoration of $sp^2$ carbon after reduction of GO.

**Table 3** Sheet resistances of inkjet-printed conductive patterns with composite Ink C and graphene inks onto 100% cotton fabrics

| Inkjet Inks and Substrates | Sheet Resistance (Ω/Sq) |
| --- | --- |
| 6L BS8 inks printed onto NP1 (12L) printed 100% Cotton | ~161.55 |
| 6L BS8 inks printed onto untreated 100% Cotton | ~2238.45 |
| 6L Composite Ink C printed onto NP1 (12L) printed 100% Cotton | ~2.11 |
| 6L Composite Ink C printed onto untreated 100% Cotton | ~30.89 |

Like previous study, here we printed 12 layers of NP1 on a rough and porous textile surface. However, we printed highly conductive graphene/silver composite (Ink C) as subsequent conductive layers (6L). Table 3 shows the sheet resistances of conductive patterns printed on untreated and 12 layer NP1 inkjet-printed hydrophobic textiles using graphene ink and Composite ink C. The sheet resistance of NP1 printed textiles with BS8 ink was found to be ~161.55 Ω/Sq and that for untreated cotton was ~2238.45 Ω/sq; which were significantly reduced to 2.11 Ω/Sq and 30.89 Ω/Sq for composite Ink C, Table 3. The inkjet printing of NP1 onto textiles produce a smooth textile surface by reducing the surface roughness; thus provides better interconnections between flakes and NPs.

SEM images of inkjet-printed e-textiles with composite Ink C (Figure 4a-d) provide further evidence of the effect of surface pre-treatment (NP1) on the conductivity of e-textiles. The inkjet printing of hydrophobic NP1 onto cotton fabrics provides interfacial interaction between fibres, Figure 4(c-d), which helped to produce a continuous film of Ag NPs and imparted very good inter-connections between graphene sheets. Therefore, the sheet resistances of the conductive patterns onto NP1 printed cotton are found to be much lower than that of untreated textiles.

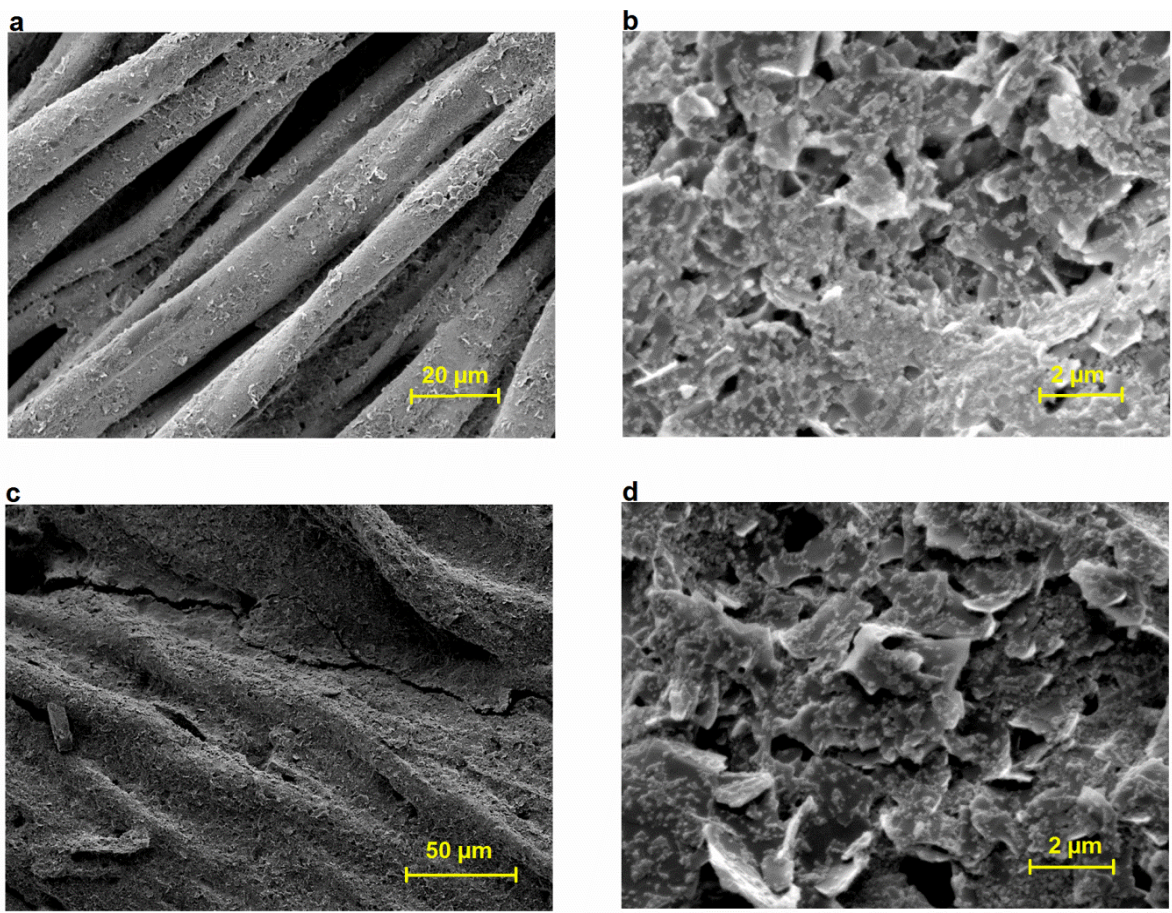

**Figure 4** SEM images of inkjet-printed composite Ink C (6L) onto untreated (top: a-b) and NP1 (12L) printed (below: c-d) 100% cotton fabrics.

We further investigated the flexibility of graphene/silver composite ink printed wearable e-textiles. Figure 5(a-b) shows almost a stable variation in the resistance up to 1000 bending-releasing and 10 folding and releasing cycles. Moreover, no visible change of printed conductive track was observed due to such mechanical actions. This demonstrates excellent flexibility of the graphene/silver composite ink device, which is in agreement with our previous study[30]. In order to demonstrate the suitability of the printed textiles for wearable strain sensor application, we mounted the printed fabric on a wrist joint and recorded the electrical resistance variation continuously during cyclic upward and downward bending of the wrist. Figure 5 (c-d) shows almost repeatable response over a period of time and the capability of the printed textiles to capture mechanical events such as bending/unbending.

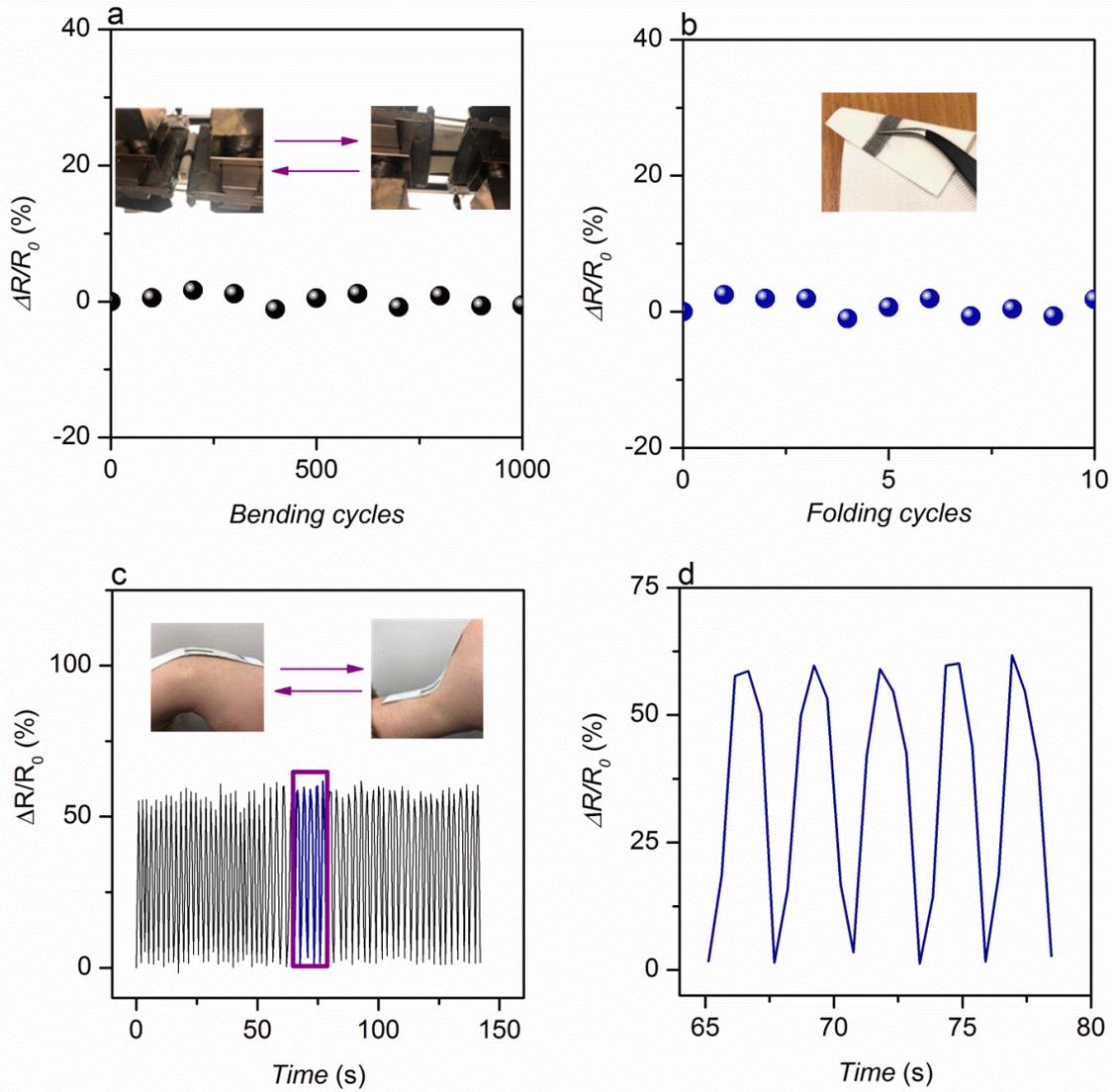

**Figure 5** Electrical resistance variation of graphene/silver composite ink printed wearable e-textiles: (a) under cyclic bending for 1000 times; (b) for 10 folding–releasing cycles; (c) with the upward and downward movements of printed e-textiles mounted on a wrist joint; (d) Expanded version of purple box in (c) from 65 s to 79 s.

**Conclusions**

We report inkjet printing of graphene-based composite inks decorated with silver nanoparticles for the first time. The conductivity of graphene-based e-textiles can be increased by layer by layer inkjet deposition of SA-ag/BS8 composite ink. The graphene/silver composites approach could potentially reduce the production cost of inkjet-printed wearable e-textiles by reducing the silver consumption. This will open up so many potential applications of graphene-based wearable e-textiles where higher electrical conductivity is desired.

**Methods**

**Materials**

Highly concentrated graphene dispersion (BS8, 8 wt.-%) was supplied by BGT Materials Limited, UK. The graphene dispersion (BS8) was stabilised using a non-ionic polymer-type surfactant. Silver nanoparticle inks (SA-Ag, 30-35 wt.-%), Triethylene glycol monomethyl ether (TEGMME), Polyvinylpyrrolidone (PVP) of 10 K molecular weight and Triton X-100 were purchased from Sigma-Aldrich. Nano 60 PEL paper was purchased from Printed Electronics Limited, UK. 100% Cotton woven fabrics (BD022) were supplied by Royal TenCate, Netherlands. Hydroxyl functionalised cross-linked styrene/divinylbenzene nanoparticles (NP1) were synthesized by following our previously reported method.[10]

**Ink Formulation and Fabrication**

In order to find the optimum percolation threshold for diluted SA-Ag ink, a series of composite inks was formulated by blending BS8, TEGMME and 1% PVP (in TEGMME) with SA-Ag inks. The formulated composite inks were deposited onto PEL paper using a triple reservoir cube film applicator (TQC, Netherland) and cured at 150 °C for 1 hr to form 90 μm thick conductive films.

Composite inkjet inks were formulated with an optimised concentration of SA-Ag inks and various concentrations of BS8 dispersion and 1 wt.-% PVP in TEGMME. The formulated inks were inkjet-printed using a Dimatix DMP-2800 inkjet printer (Fujifilm Dimatix Inc., Santa Clara, USA) which can create and define patterns over an area of 200 × 300 mm and handle substrates up to 25 mm thick, being adjustable in the Z direction. This printer is equipped with a disposable piezo inkjet cartridge and the nozzle plate consists of a single row of 16 nozzles of 21.5 μm diameter spaced 254 μm with typical drop diameter 27 μm and 10 pl drop size. Print head height was adjusted to 0.75 mm and the formulated inks were jetted at 37 °C temperature, using frequent cleaning cycles during the printing. A few layers (1-5L) of composite inks were printed to produce a conductive pattern of 1 cm$^2$ area and thermally-cured at 150 °C for 1 hr in an oven to sinter the conductive inks.

In order to demonstrate the potential electronic textiles applications of graphene-based composite inks, a hydrophobic coating was inkjet-printed onto 100% cotton plain twill fabrics (B022) by depositing 12 layers of nanoparticles (NP1) as reported in our previous study[10]. Subsequently, six layers (6L) of graphene inks (formulated from BS8 dispersion) or composite inks C were inkjet-printed onto hydrophobic areas of cotton fabrics.

**Characterisation**

The Raman spectra of BS8 was obtained from a low power (<1 mW) He-Ne laser (1.96 eV, 633 nm) in Renishaw 2000 spectrometer. The viscosity of formulated inks was measured using a Brookfield DV-II+PRO programmable digital viscometer at 25 °C temperature and surface tension was measured by using a torsion balance (model OS) for surface and interfacial tension measurement. Thermogravimetric Analysis (TGA) was conducted to investigate the thermal stability of formulated inks using a TGA Q500 (TA Instruments, USA). A Philips XL 30 Field Emission Gun Scanning Electron Microscope (SEM) was used to analyse the surface topography. Printed samples were gold-palladium (Au-Pd) coated for 90 seconds and assessed under FEG SEM with the following operating parameters: 6.0 KV, spot size 2.0, 10 mm WD and magnification: X500 to X40000. A Jandel four-point probe system (Jandel Engineering Ltd, Leighton, UK) was employed to measure the sheet resistance of the conductive patterns. The sheet resistance was calculated from the average of six measurements and multiplied by a correction factor of ~4.5324.

A Zwick/Roell tensile tester (Zwick Roell Group, Germany) was used to measure the change of the resistance during bending and unbending of graphene/silver composites ink printed textiles. A National Instrument 9219 data acquisition card (NI, American) was used to record the change of the resistance during bending, folding and upward-downward movement of wrist joint.

**Acknowledgements**

The authors kindly acknowledge the financial support from The University of Manchester Research Impact Scholarship and the government of Bangladesh for the PhD study of Nazmul karim and Shaila Afroj, respectively. Authors also acknowledges the funding from EU Graphene Flagship Program, European Research Council Synergy Grant Hetero2D, the Royal Society, and Engineering and Physical Sciences Research Council, U.K. (Grant Number: EP/N010345/1, 2015).

**References**

1. Kim, D. H. *et al.* Epidermal Electronics. *Science* **333**, 838-843 doi:10.1126/science.1206157 (2011).
2. Matsuhisa, N. *et al.* Printable Elastic Conductors with a High Conductivity for Electronic Textile Applications. *Nat. Commun.* **6**, 7461, doi:10.1038/ncomms8461 (2015).
3. Abdelkader, A. M. *et al.* Ultraflexible and Robust Graphene Supercapacitors Printed on Textiles for Wearable Electronics Applications. *2D Mater.* **4**, 035016 (2017).
4. Jung, M. *et al.* All-Printed and Roll-to-Roll-Printable 13.56-MHz-Operated 1-bit RF Tag on Plastic Foils. *IEEE Trans. Electron Devices* **57**, 571-580 (2010).
5. Zhou, L. *et al.* All-Organic Active Matrix Flexible Display. *Appl. Phys. Lett.* **88**, 083502, doi:10.1063/1.2178213 (2006).
6. Gelinck, G. H. *et al.* Flexible Active-Matrix Displays and Shift Registers Based on Solution-Processed Organic Transistors. *Nat. Mater.* **3**, 106, doi:10.1038/nmat1061 (2004).
7. Kulkarni, M. V., Apte, S. K., Naik, S. D., Ambekar, J. D. & Kale, B. B. Ink-Jet Printed Conducting Polyaniline Based Flexible Humidity Sensor. *Sensors and Actuators B: Chemical* **178**, 140-143, doi:https://doi.org/10.1016/j.snb.2012.12.046 (2013).
8. Matsuhisa, N. *et al.* Printable Elastic Conductors by In Situ Formation of Silver Nanoparticles from Silver Flakes. *Nat. Mater.* **16**, 834, doi:10.1038/nmat4904 (2017).
9. Chauraya, A. *et al.* Inkjet Printed Dipole Antennas on Textiles for Wearable Communications. *IET Microwaves, Antennas & Propagation* **7**, 760-767 (2013).
10. Karim, N. *et al.* All Inkjet-Printed Graphene-Based Conductive Patterns for Wearable E-Textile Applications. *J. Mater. Chem. C* **5**, 11640-11648, doi:10.1039/C7TC03669H (2017).
11. Stoppa, M. & Chiolerio, A. Wearable Electronics and Smart Textiles: A Critical Review. *Sensors (Basel, Switzerland)* **14**, 11957-11992, doi:10.3390/s140711957 (2014).
12. Secor, E. B., Prabhumirashi, P. L., Puntambekar, K., Geier, M. L. & Hersam, M. C. Inkjet Printing of High Conductivity, Flexible Graphene Patterns. *J. Phys. Chem. Lett.* **4**, 1347-1351, doi:10.1021/jz400644c (2013).
13. Karim, M. N., Afroj, S., Rigout, M., Yeates, S. G. & Carr, C. Towards UV-Curable Inkjet Printing of Biodegradable Poly (Lactic Acid) Fabrics. *J. Mater. Sci.* **50**, 4576-4585 (2015).
14. Kamyshny, A. & Magdassi, S. Conductive Nanomaterials for Printed Electronics. *Small* **10**, 3515-3535, doi:doi:10.1002/smll.201303000 (2014).
15. Xu, L. Y., Yang, G. Y., Jing, H. Y., Wei, J. & Han, Y. D. Ag–Graphene Hybrid Conductive Ink for Writing Electronics. *Nanotechnology* **25**, 055201 (2014).
16. Son, D. *et al.* An Integrated Self-Healable Electronic Skin System Fabricated via Dynamic Reconstruction of A Nanostructured Conducting Network. *Nat. Nanotechnol.* **13**, 1057 (2018).
17. Ma, R., Suh, D., Kim, J., Chung, J. & Baik, S. A Drastic Reduction in Silver Concentration of Metallic Ink by the Use of Single-Walled Carbon Nanotubes Decorated with Silver Nanoparticles. *J. Mater. Chem.* **21**, 7070-7073, doi:10.1039/C1JM10850F (2011).
18. Russo, A. *et al.* Pen-on-Paper Flexible Electronics. *Adv. Mater.* **23**, 3426-3430, doi:doi:10.1002/adma.201101328 (2011).
19. Novoselov, K. S. *et al.* Electric Field Effect in Atomically Thin Carbon Films. *Science* **306**, 666-669 (2004).
20. Novoselov, K. *et al.* Two-Dimensional Atomic Crystals. *Proc. Natl. Acad. Sci.* **102**, 10451-10453 (2005).
21. Karim, N. *et al.* Graphene-Based Surface Heater for De-Icing Applications. *RSC Adv.* **8**, 16815-16823, doi:10.1039/C8RA02567C (2018).
22. Sarker, F. *et al.* High-Performance Graphene-Based Natural Fiber Composites. *ACS Appl. Mater. Interfaces* **10**, 34502-34512, doi:10.1021/acsami.8b13018 (2018).

23  Sarker, F. *et al.* Ultra-High Performance of Nano-Engineered Graphene-Based Natural Jute Fiber Composites. *ACS Appl. Mater. Interfaces* **In Press** (2019).
24  Li, J. *et al.* Efficient Inkjet Printing of Graphene. *Adv. Mater.* **25**, 3985-3992 (2013).
25  Dua, V. *et al.* All-Organic Vapor Sensor Using Inkjet-Printed Reduced Graphene Oxide. *Angew. Chem.* **122**, 2200-2203 (2010).
26  Le, L. T., Ervin, M. H., Qiu, H., Fuchs, B. E. & Lee, W. Y. Graphene Supercapacitor Electrodes Fabricated by Inkjet Printing and Thermal Reduction of Graphene Oxide. *Electrochem. Commun.* **13**, 355-358 (2011).
27  Huang, L., Huang, Y., Liang, J., Wan, X. & Chen, Y. Graphene-Based Conducting Inks for Direct Inkjet Printing of Flexible Conductive Patterns and Their Applications in Electric Circuits and Chemical Sensors. *Nano Res.* **4**, 675-684 (2011).
28  Eda, G. & Chhowalla, M. Chemically Derived Graphene Oxide: Towards Large-Area Thin-Film Electronics and Optoelectronics. *Adv. Mater.* **22**, 2392-2415 (2010).
29  Karim, N. *et al.* Scalable Production of Graphene-Based Wearable E-Textiles. *ACS Nano* **11**, 12266-12275, doi:10.1021/acsnano.7b05921 (2017).
30  Afroj, S. *et al.* Engineering Graphene Flakes for Wearable Textile Sensors via Highly Scalable and Ultrafast Yarn Dyeing Technique. *ACS Nano*, doi:10.1021/acsnano.9b00319 (2019).
31  Beach, C., Karim, N. & Casson, A. J. Performance of Graphene ECG Electrodes Under Varying Conditions. in *2018 40th Annual International Conference of the IEEE Engineering in Medicine and Biology Society (EMBC).*  3813-3816.
32  Zhao, J., Pei, S., Ren, W., Gao, L. & Cheng, H. M. Efficient Preparation of Large-Area Graphene Oxide Sheets for Transparent Conductive Films. *ACS nano* **4**, 5245-5252 (2010).
33  Torrisi, F. *et al.* Inkjet-Printed Graphene Electronics. *ACS Nano* **6**, 2992-3006 (2012).
34  Han, X. *et al.* Scalable, Printable, Surfactant-Free Graphene Ink Directly from Graphite. *Nanotechnology* **24**, 205304 (2013).
35  Gao, Y., Shi, W., Wang, W., Leng, Y. & Zhao, Y. Inkjet Printing Patterns of Highly Conductive Pristine Graphene on Flexible Substrates. *Ind. Eng. Chem. Res.* **53**, 16777-16784 (2014).
36  Chun, K. Y. *et al.* Highly Conductive, Printable and Stretchable Composite Films of Carbon Nanotubes and Silver. *Nat. Nanotechnol.* **5**, 853 (2010).
37  Yang, J., Zang, C., Sun, L., Zhao, N. & Cheng, X. Synthesis of Graphene/Ag Nanocomposite with Good Dispersibility and Electroconductibility via Solvothermal Method. *Mater. Chem. Phys.* **129**, 270-274 (2011).
38  Wang, G. *et al.* Annealed Graphene Sheets Decorated with Silver Nanoparticles for Inkjet Printing. *Chem. Eng. J.* **260**, 582-589, doi:https://doi.org/10.1016/j.cej.2014.09.037 (2015).
39  Jie, Z. *et al.* Microstructure-Tunable Highly Conductive Graphene–Metal Composites Achieved by Inkjet Printing and Low Temperature Annealing. *J Micromech Microeng* **28**, 035006 (2018).
40  Bourlinos, A. B. *et al.* Aqueous-Phase Exfoliation of Graphite in the Presence Of Polyvinylpyrrolidone for the Production of Water-Soluble Graphenes. *Solid State Commun.* **149**, 2172-2176, doi:https://doi.org/10.1016/j.ssc.2009.09.018 (2009).
41  Knoerr, M. & Schletz, A. Power Semiconductor Joining Through Sintering of Silver Nanoparticles: Evaluation of Influence of Parameters Time, Temperature and Pressure on Density, Strength and Reliability. in *2010 6th International Conference on Integrated Power Electronics Systems.*  1-6.
42  Zhang, L. L., Zhou, R. & Zhao, X. S. Graphene-Based Materials As Supercapacitor Electrodes. *J. Mater. Chem.* **20**, 5983-5992, doi:10.1039/C000417K (2010).
43  Dua, V. *et al.* All-Organic Vapor Sensor Using Inkjet-Printed Reduced Graphene Oxide. *Angew. Chem. Int. Ed.*  **49**, 2154-2157, doi:doi:10.1002/anie.200905089 (2010).
44  Hu, G. *et al.* Functional Inks and Printing of Two-Dimensional Materials. *Chem. Soc. Rev.* **47**, 3265-3300, doi:10.1039/C8CS00084K (2018).

| | |
|---|---|
| 45 | Wajid, A. S. *et al.* Polymer-Stabilized Graphene Dispersions at High Concentrations in Organic Solvents for Composite Production. *Carbon* **50**, 526-534, doi:https://doi.org/10.1016/j.carbon.2011.09.008 (2012). |
| 46 | Teranishi, T. & Miyake, M. Size Control of Palladium Nanoparticles and Their Crystal Structures. *Chem. Mater.* **10**, 594-600, doi:10.1021/cm9705808 (1998). |
| 47 | Yang, J., Zang, C., Sun, L., Zhao, N. & Cheng, X. Synthesis of Graphene/Ag Nanocomposite with Good Dispersibility and Electroconductibility via Solvothermal Method. *Mater. Chem. Phys.* **129**, 270-274, doi:https://doi.org/10.1016/j.matchemphys.2011.04.002 (2011). |
| 48 | Salvado, R., Loss, C., Gonçalves, R. & Pinho, P. Textile Materials for the Design of Wearable Antennas: A Survey. *Sensors* **12**, 15841 (2012). |

## Authors Contributions

Nazmul Karim designed and prepared the experiments, measured and analysed the experimental data and drafted the manuscript. Shaila Afroj participated in measurements and drafted the manuscript. Sirui Tan carried out flexibility measurements of the printed device and its application as a strain sensor. Stephen G Yeates and Kostya S Novoselov supervised the project and contributed in drafting the manuscript.

## Competing Interests

The authors declare no competing interests.

## Corresponding Author

All Correspondence to Nazmul Karim (mdnazmul.karim@manchester.ac.uk)